\documentclass[journal=jacsat,manuscript=article]{achemso}
\usepackage[version=3]{mhchem} % Formula subscripts using \ce{}
\usepackage[]{natbib}
\usepackage{amsmath}
\usepackage{amssymb}
\usepackage{color}
\usepackage{graphicx}
\graphicspath{{figures/}}
\usepackage[english]{babel}
\usepackage{titlesec}
\usepackage{hyperref}

\author{Meysam Bagheri}
\email{meysam.bagheri@fau.de}
\author{Sudeshna Roy}
\author{Thorsten P\"oschel}
\affiliation{{Institute for Multiscale Simulation, Friedrich-Alexander-Universit\"at Erlangen-N\"urnberg, Erlangen,  Germany}}

\title[Force and the Surface Area of a Liquid Bridge] {Approximate Expressions for the Capillary Force and the Surface Area of a Liquid Bridge between Identical Spheres}

\keywords{American Chemical Society, \LaTeX}

\begin{document}
\begin{tocentry}
\centerline{\includegraphics[width=8cm,bb=195 50 900 352, clip=]{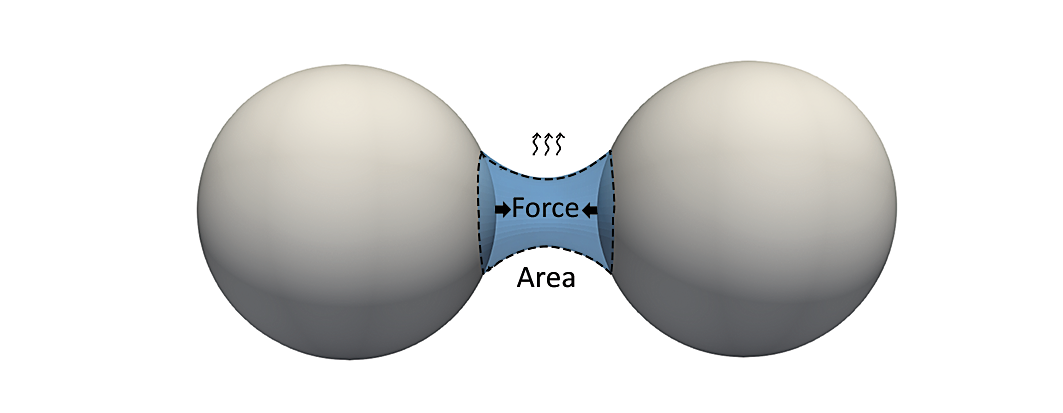}}
\end{tocentry}

\begin{abstract}
  We consider a liquid bridge between two identical spheres and provide approximate expressions for the capillary force and the exposed surface area of the liquid bridge as functions of the liquid bridge's total volume and the sphere separation distance. The radius of the spheres and the solid-liquid contact angle are parameters that enter the expressions. These expressions are needed for efficient numerical simulations of drying suspensions.
\end{abstract}

\section{Introduction}
In a certain approximation, drying suspensions can be modeled as identical spheres that are pairwise connected by liquid bridges exerting forces on the particles \cite{fustin2004,nishimoto2007,gohering2009,goehring2015desiccation, Cai2021}. The liquid bridges change over time due to diffusion, thus, the forces are time-dependent, resulting in complex phenomena such as fragmentation \cite{Zhou2006, Goehring2010, xiaolei2019, PAUL2023}. If we consider the radius of the spheres, $R$, and the solid-liquid contact angle, $\theta$, as invariant material parameters, the shape of a liquid bridge of a given volume, $V$, spanning between the spheres of separation distance $S_d$ is provided by the solution of the Young-Laplace equation \cite{Erle1971}. Therefore, the total area of the exposed surface of a liquid bridge, $A$, and the capillary force are functions of $V$ and $S_d$. The evolution of the capillary force is thus determined by the evaporation rate, which in turn depends on $A$.

For the numerical simulation of macroscopic systems of drying suspension with millions of liquid bridges, we need an efficient way to compute the capillary force exerted by a liquid bridge and its exposed surface area as functions of the liquid volume and the distance of the spheres. While the straightforward computation of these quantities requires the numerical solution of a partial differential equation, the current paper presents corresponding approximations suitable for Discrete Element Method (DEM) and Molecular Dynamics (MD) simulations.

Our approach is based on the numerical solution of the Young-Laplace equation.  Several approximate solutions of the Young-Laplace equation have been described in the literature. These approximations aim to simplify the mathematical representation of the shape of the liquid bridge \cite{Willett2000, Fisher1926, Derjaguin1934, Butt2009, Rabinovich2005, Gladky2014}. However, none of them facilitates the forward calculation of the capillary force and the free surface of the bridge as functions of the liquid volume and the distance between the surfaces the liquid bridge spans.

\section{Problem Description}

Consider the situation sketched in \autoref{fig LB}:
\begin{figure}[!htb]
  \centering
  \includegraphics[width=9.7cm]{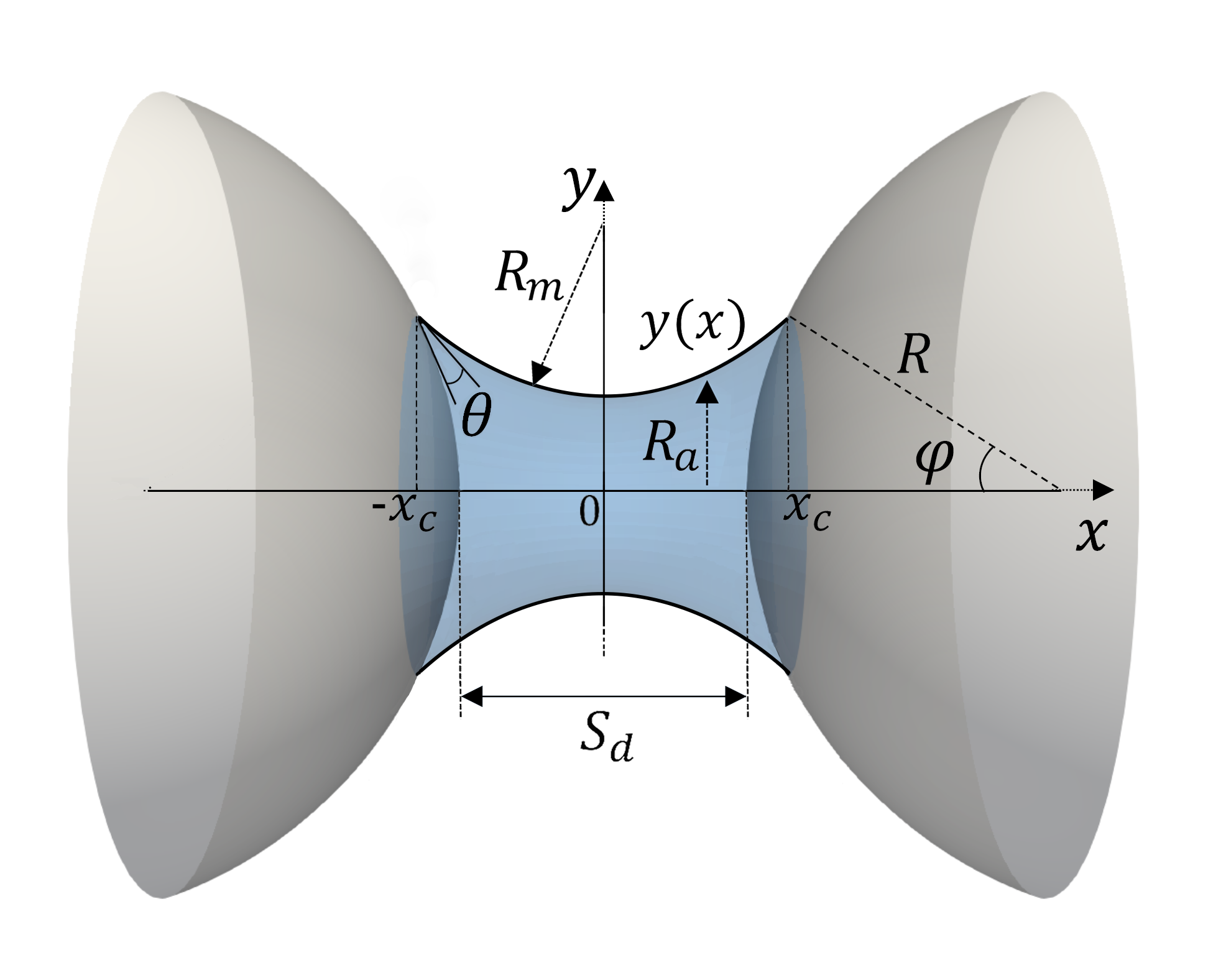}
  \caption{The shape of a liquid bridge between two identical spheres of radius $R$ and distance $S_d$ is a solution of the Young-Laplace equation with the solid-liquid contact angle, $\theta$, the half-filling angle, $\varphi$, the surface tension coefficient, $\gamma$. $R_m$ and $R_a$ are meridian and azimuth radii, respectively.}
  \label{fig LB}
\end{figure}
A liquid bridge spans between two spheres of radius $R$ and distance $S_d$. Liquid bridges between spheres can be attractive or repulsive, depending on the total volume and the boundary conditions. For application in drying suspension, attractive forces are relevant. In this case, the analysis simplifies since surface tension enforces an axisymmetric shape of the liquid bride, as sketched in \autoref{fig LB}. This is different for the bulbous shape of the liquid bridge (large volume) where the static solution can be asymmetric \cite{FarmerBird:2015}. This case is not considered here. Also, the influence of gravity shall be neglected. 

The force exerted by the liquid bridge on the spheres results from two contributions: the $x$-component of the surface tension and the $x$-component of the hydrostatic pressure acting on the liquid bridge. For the evaluation of the force, we are free to choose any $x$-position, however, without knowing the shape of the liquid bridge, the projections can be evaluated only at the three-phase contact, $x=\pm x_c$. Here, we have the force
\begin{equation}
  \label{eq:Fgamma}
  F_\gamma = \gamma\,2\pi R \sin \varphi\,
\end{equation}
acting tangential to the bridge, where $\varphi$ is the half-filling angle.  Its $x$-component reads
\begin{equation}
  \label{eq:Fgamma_x}
  F_\gamma^x = 2\pi\gamma R \sin \varphi \sin (\varphi+\theta)\,,
\end{equation}
with the fluid-solid contact angle $\theta$. The $x$-component of the force due to  the hydrostatic pressure reads
\begin{equation}
  \label{eq:Fpress_x}
  F_{\Delta p}^x = \pi R^2\Delta p \sin^2\varphi\,, 
\end{equation}
where $\Delta p$ is the pressure drop at the air-liquid interface. We relate these forces to the characteristic force, $\gamma R$, and obtain the  total dimensionless force
\begin{equation}
  \label{eq:force1}
  F^*\equiv \frac{1}{\gamma R}\left( F_\gamma^x+F_{\Delta p}^x\right)= 2\pi\sin \varphi \sin (\varphi+\theta) + {\pi}\sin^2\varphi \frac{\Delta p\,R}{\gamma}\,.
\end{equation}
The superscript "*" refers to dimensionless quantities here and in the following. The pressure drop follows from the Young-Laplace equation, which relates the pressure difference to the shape of the surface, characterized by its mean curvature, $H$:
\begin{equation}
  \label{eq:YLgen}
  \Delta p =  \gamma H\,.
\end{equation}
Therefore,
\begin{equation}
  \label{eq:H}
  H^*\equiv\frac{H}{\frac{1}{R}}=\frac{\Delta p\,R}{\gamma}\,.
\end{equation}
Equation \eqref{eq:force1}  for  the dimensionless force between the spheres reads then
\begin{equation}
  \label{eq:force}
   F^* = 2\pi\sin \varphi \sin (\varphi+\theta) + {\pi}\sin^2\varphi H^*\,.
\end{equation}

For the axisymmetric liquid bridge, $H$ can be expressed through the radial extension of the bridge, $y(x)$,  see \autoref{fig LB}  \cite{lian1993}
\begin{equation}
  \label{Young-Laplace}
  H=
  \frac{
    \frac{\text{d}^2 y(x)} {\text{d} x^2}
  }{
    \left[ 1+\left(\frac{\text{d}y(x)}{\text{d} x}\right)^2\right]^{3/2}
  }
  -\frac{1}{y(x)\sqrt{1+\left(\frac{\text{d}y(x)}{\text{d} x}\right)^2}}\,.
\end{equation}
The boundary conditions 
\begin{equation}
  \label{boundary conditions}
      \begin{split}
          \left.\frac{\text{d}y(x)}{\text{d}x}\right|_{x=\pm x_c} &= R\sin{\varphi}\\
          \left.\frac{{\text{d}^2y(x)}}{\text{d}x^2}\right|_{x=\pm x_c} &= \pm\cot{(\varphi+\theta)}
      \end{split}
\end{equation}
are given at the three-phase contact points,  $x=\pm x_c$, with  
\begin{equation}
\label{eq:xc}
    x_c\equiv \frac{S_d}{2}+R-R\,\cos(\varphi)\,.
\end{equation}
Introducing the dimensionless coordinates, $x^*_c\equiv x_c/R$ and $y^*(x)\equiv y(x)/R$,  from the solution $y(x)$ of Eq. \eqref{Young-Laplace}, we obtain the dimensionless volume of the liquid bridge, $V^*\equiv V/R^3$,
\begin{equation}
  \label{eq:v}
  V^*=\int_{-x^*_c}^{x^*_c}  \pi \left[y^*(x)\right]^2 \,\text{d}x^* - \frac{2\pi}{3} (1-\cos{\varphi})^2(2+\cos{\varphi})\,.
\end{equation}
The second term on the right-hand side is the volume of the two wet caps on each particle, which must be subtracted from the integral to obtain the liquid volume.

The dimensionless free surface area of the liquid bridge, $A^*\equiv A/R^2$, is
\begin{equation}
  \label{eq:a}
    A^*=\int_{-x^*_c}^{x^*_c} 2\pi y^*(x)\sqrt{1+ \left(\frac{\text{d} y^*(x)}{\text{d}x^*}\right)^2} \,\text{d}x^*\, .
\end{equation}

Equations \eqref{Young-Laplace}, \eqref{eq:force}, \eqref{eq:v}, and \eqref{eq:a} establish a  mathematical system that relates the force, $F^*$, the volume, $V^*$, and the accessible surface area, $A^*$, of a liquid bridge to its profile. The coefficient of surface tension, $\gamma$, the contact angle, $\theta$, and the separation distance between the spheres, $S^*_d \equiv S_d/R$, enter as physical parameters. The radius, $R$, of the spheres does not enter as all quantities are adimensionalized using $R$ as the characteristic length.

From this set of equations, in the following, we derive approximative expressions, $F^*(S^*_d,V^*)$ and $A^*(S^*_d,V^*)$, for use in efficient numerical simulations.

\section{Results and discussion}
\label{ResultsAndDiscussion}
\subsection{Approximative expression $A^*(S^*_d,V^*)$ for the accessible surface}
\label{sec:ApproxSurface}
\subsubsection{Outline}
Using the numerical scheme detailed in \ref{sec:NumMethod},  we solve numerically the Young-Laplace equation \eqref{Young-Laplace}, where the scaled volume, $V^*$, enters as a parameter. From the numerical results, step-by-step, we develop an approximation formula for the surface area of the liquid bridge as a function of the liquid volume, $V^*$, and the distance between the spheres, $S_d^*$. The contact angle, $\theta$, is the single parameter of the formula. In  Sec. \ref{sec:Sd0_theta0} we start with the case $S_d^*=0$ and $\theta=0$ to obtain $A^*\left.(S^*_d,V^*)\right|_{S_d^*=0;\, \theta=0}$. In Sec. \ref{sec:Sd0} we release the constraint $\theta=0$ to find $\left.A^*(S^*_d,V^*)\right|_{S_d^*=0}$. Eventually, in Sec. \ref{sec:A} , we also release the condition $S^*_d=0$ to obtain our final result, $A^*(S^*_d,V^*)$.

\subsubsection{Surface area as a function of the volume for particle separation $S^*_d=0$ and contact angle $\theta=0$}
\label{sec:Sd0_theta0}

\begin{figure}[!htb]
    \centering
    \includegraphics[width=16.5cm]{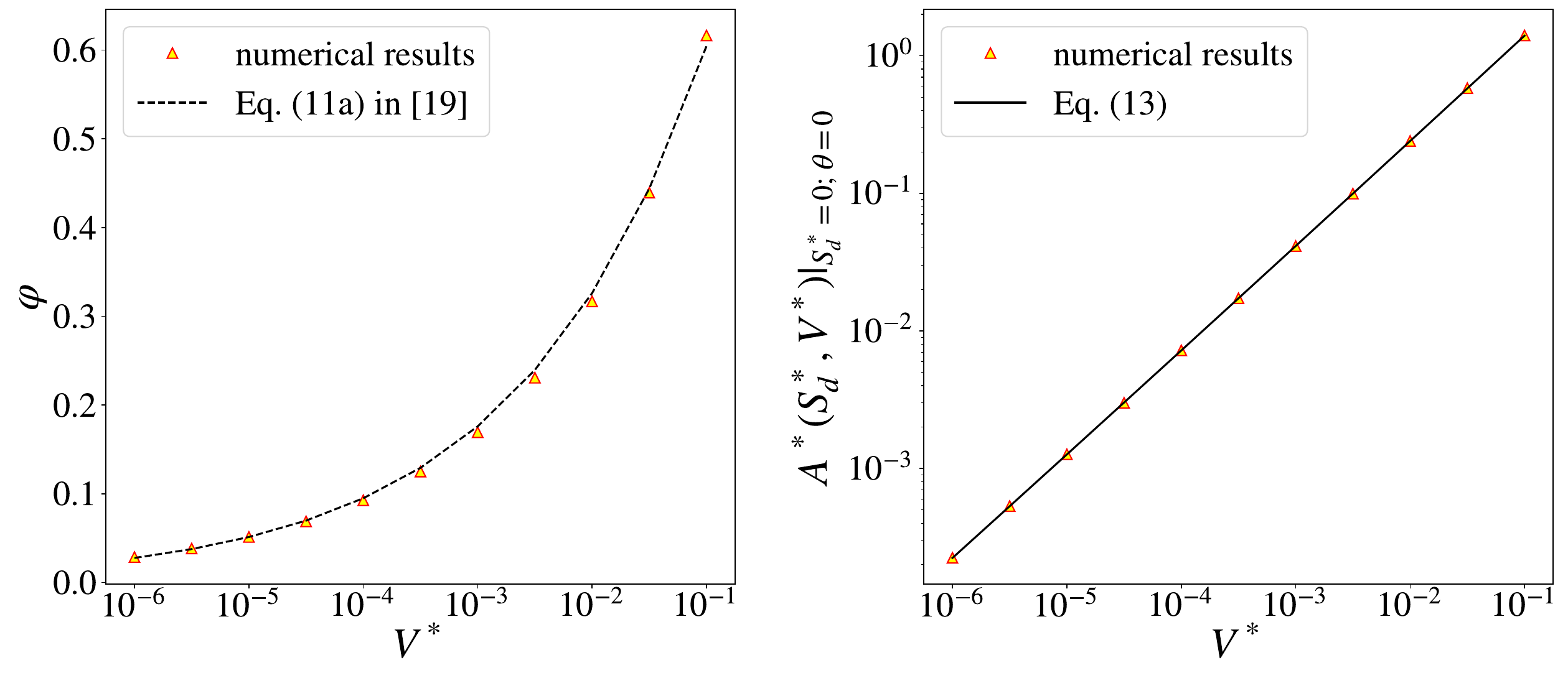}
 \caption{Numerical results for $S_d^*=0$ and $\theta=0$.  (a) Half-filling angle as a function of the volume, $\varphi(V^*)$ together with the approximation Eq. (11a) in Ref. \citenum{Lian2016}.    (b) The surface area of the liquid bridge as a function of the volume, $\left.A^*(S_d^*,V^*)\right|_{S_d^*=0;\, \theta=0}$, and the 
 }fit formula Eq. \eqref{eq area vs volume}
\label{fig:half-filling-angle}
\end{figure}

We start with the special case of vanishing separation distance, $S^*_d=0$, and contact angle, $\theta=0$. Figure \ref{fig:half-filling-angle}a shows the half-filling angle, $\varphi$, as a function of the volume, $V^*$. 
The numerical data points agree perfectly with the results by \citeauthor{Lian2016}\cite{Lian2016}, supporting the validity of our simulation method.

The surface area as a function of volume follows a power law with a small correction, see \autoref{fig:half-filling-angle}b. The best fit up to the second order in $\ln V^*$ is 
\begin{equation}
    \label{eq area vs volume}
 \ln \left.A^*(S_d^*,V^*)\right|_{S_d^*=0;\, \theta=0} 
 =0.000734 (\ln{V^*})^2 + 0.771\ln{V^*}+2.107.
\end{equation}

\begin{figure}[!htb]
    \centering
    \includegraphics[width=10cm]{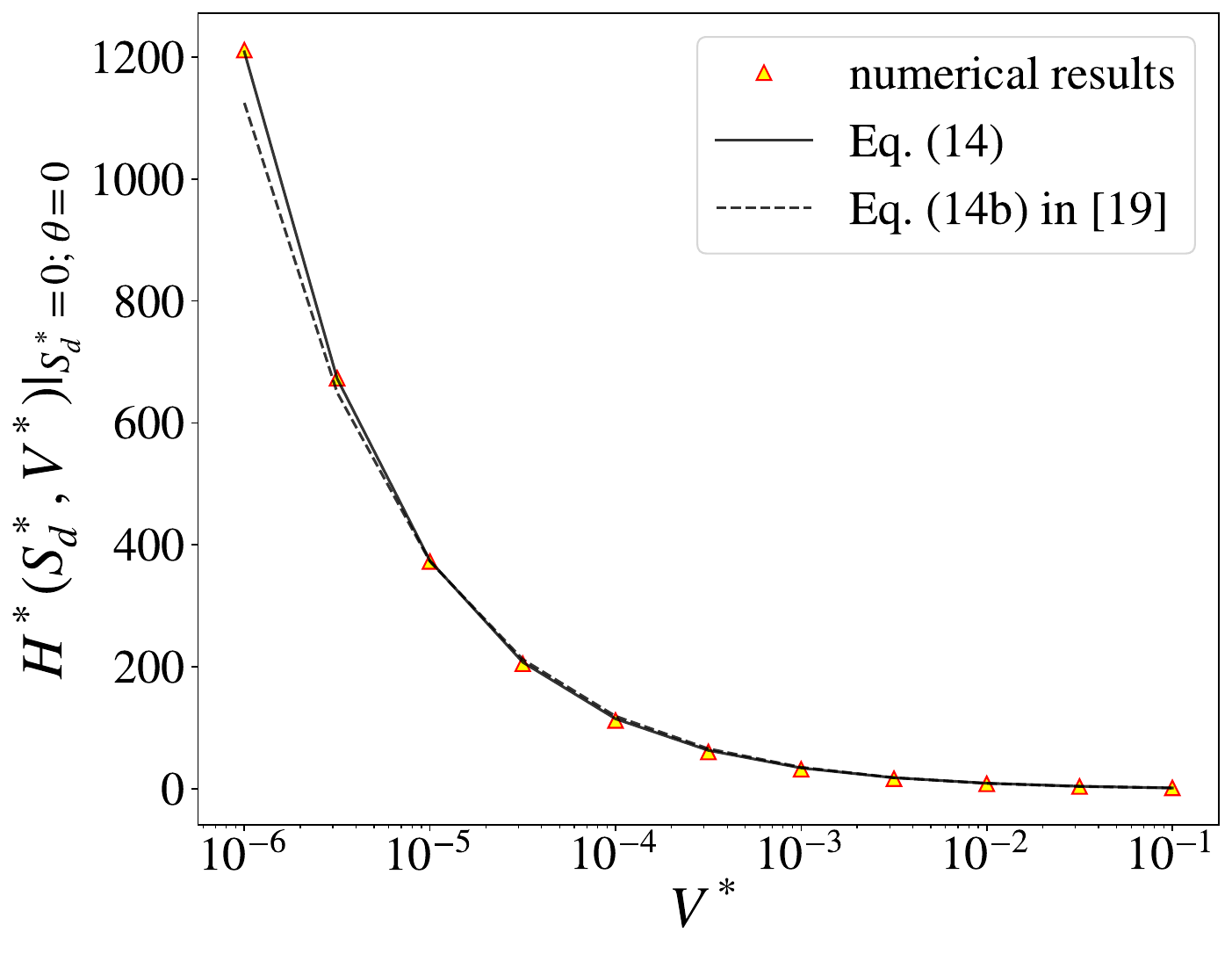}
 \caption{Dimensionless mean curvature as a function of the volume,
   $\left.H^*(S_d^*,V^*)\right|_{S_d^*=0;\, \theta=0}$. The lines show
   the fit formula, Eq. \eqref{eq H0 fit}, and the fit by  \citeauthor{Lian2016}\cite{Lian2016}}
 \label{fig zero}
\end{figure}

As a side product, we also obtain the dimensionless mean curvature as a function of the volume, 
$\left.H^*(S_d^*,V^*)\right|_{S_d^*=0;\, \theta=0}$, 
shown in \autoref{fig zero}, 
together with the fit
\begin{equation}
\left.H^*(S_d^*,V^*)\right|_{S_d^*=0;\, \theta=0}
=1.09 \ln V^{*\,-0.5076} -2.25
    \label{eq H0 fit}
\end{equation}
which slightly improves the fit in Ref.\citenum{Lian2016} and is, moreover, algebraically simpler. 

\subsubsection{Surface area as a function of the volume for particle separation $S^*_d=0$}
\label{sec:Sd0}

\begin{figure}[!htb]
    \centering
    \includegraphics[width=10cm]{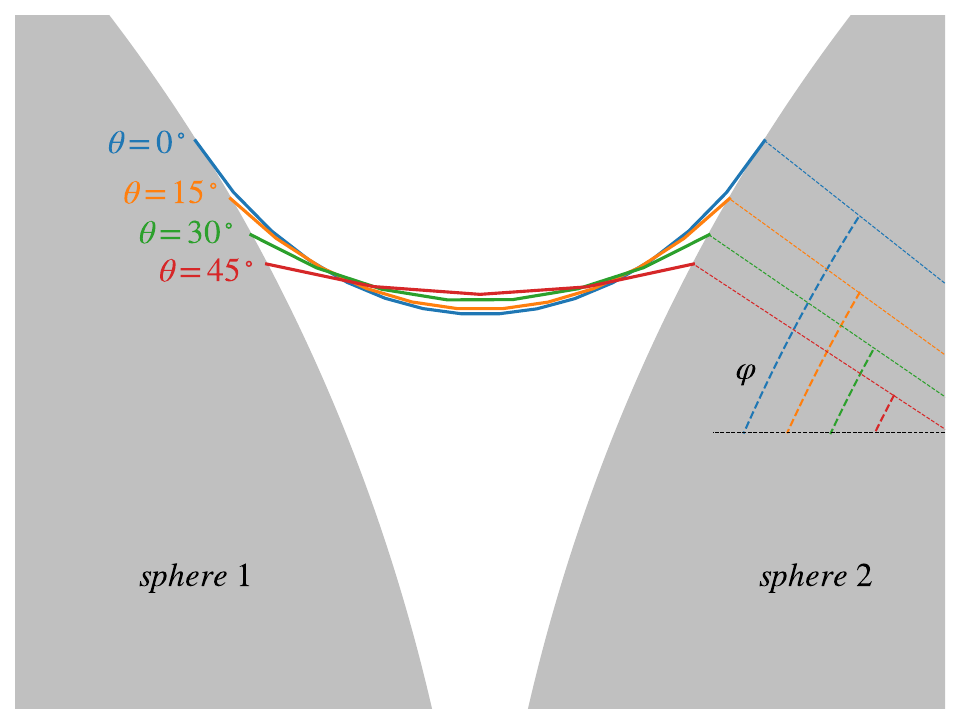}
    \caption{Profiles of a liquid bridge of volume  $V^*=0.1$, spanning between two spheres in contact ($S_d^*=0$). When keeping the volume invariant, varying the contact angle $\theta\in\{0, 15^\circ, 30^\circ, 45^\circ\}$ leads to corresponding changes in the half-filling angle, $\varphi$, and the mean curvature, $\left. H^*\right|_{S^*d=0}$. The spheres are partially shown in grey.}
    \label{fig liquid bridge profile}
\end{figure}

\begin{figure}[!htb]
    \centering
    \includegraphics[width=16.5cm]{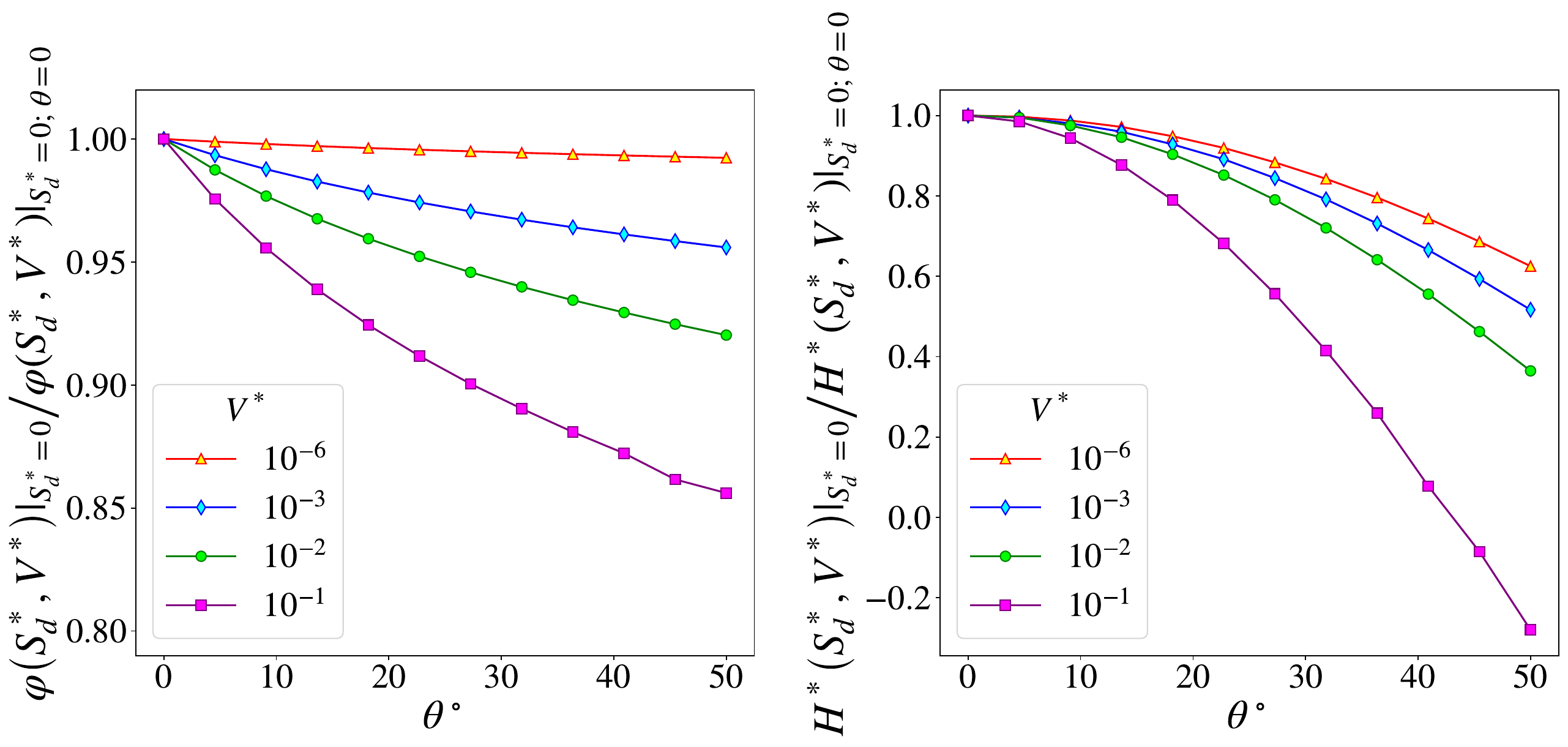}
    \caption{Numerical results for liquid bridges between spheres in contact, $S^*_d=0$, and constant volume, $V^*\in\left\{10^{-6}, 10^{-3}, 10^{-2}, 10^{-1}\right\}$. (a) normalized half-filling angle as a function of the contact angle, 
    $\left.\left. \varphi\left(S_d^*,V^*\right)\right|_{S^*_d=0} \right/ \left.\varphi(S_d^*,V^*)\right|_{S^*_d=0;\, \theta=0}$.   
    (b) normalized mean curvature as a function of the contact angle,
    $\left.\left. H^*\left(S_d^*,V^*\right)\right|_{S^*_d=0} \right/ \left.H^*(S_d^*,V^*)\right|_{S^*_d=0;\, \theta=0}$
    .}
    \label{fig PhiAndH/ContactAngle}
\end{figure}

When changing the contact angle, $\theta$, keeping the liquid volume fixed, the half-filling angle, $\varphi$, and the mean curvature, $H^*$, change correspondingly, see \autoref{fig liquid bridge profile}.
Figure  \ref{fig PhiAndH/ContactAngle}  quantifies this effect. It shows the half-filling angle and the mean curvature as functions of the contact angle, 
$\left.\varphi\left(S_d^*,V^*\right)\right|_{S^*_d=0}$ 
and
$\left.H^*\left(S_d^*,V^*\right)\right|_{S^*_d=0}$, normalized by 
$\left.\varphi(S_d^*,V^*)\right|_{S^*_d=0;\, \theta=0}$ and
$ \left.H^*(S_d^*,V^*)\right|_{S^*_d=0;\, \theta=0}$, respectively, for four values of the volume.

For the largest volume shown here, $V^*=0.1$,  the mean curvature assumes negative values for the contact angle $\theta\gtrsim 40^\circ$, indicating the pressure inside the liquid bridge exceeds the ambient pressure. The capillary pressure yields, thus, a repulsive contribution to the force between the spheres. The total force remains, however, attractive due to the dominating contribution of surface tension.\cite{Lian2016}

\begin{figure}[!htb]
    \centering
    \includegraphics[width=9.5cm]{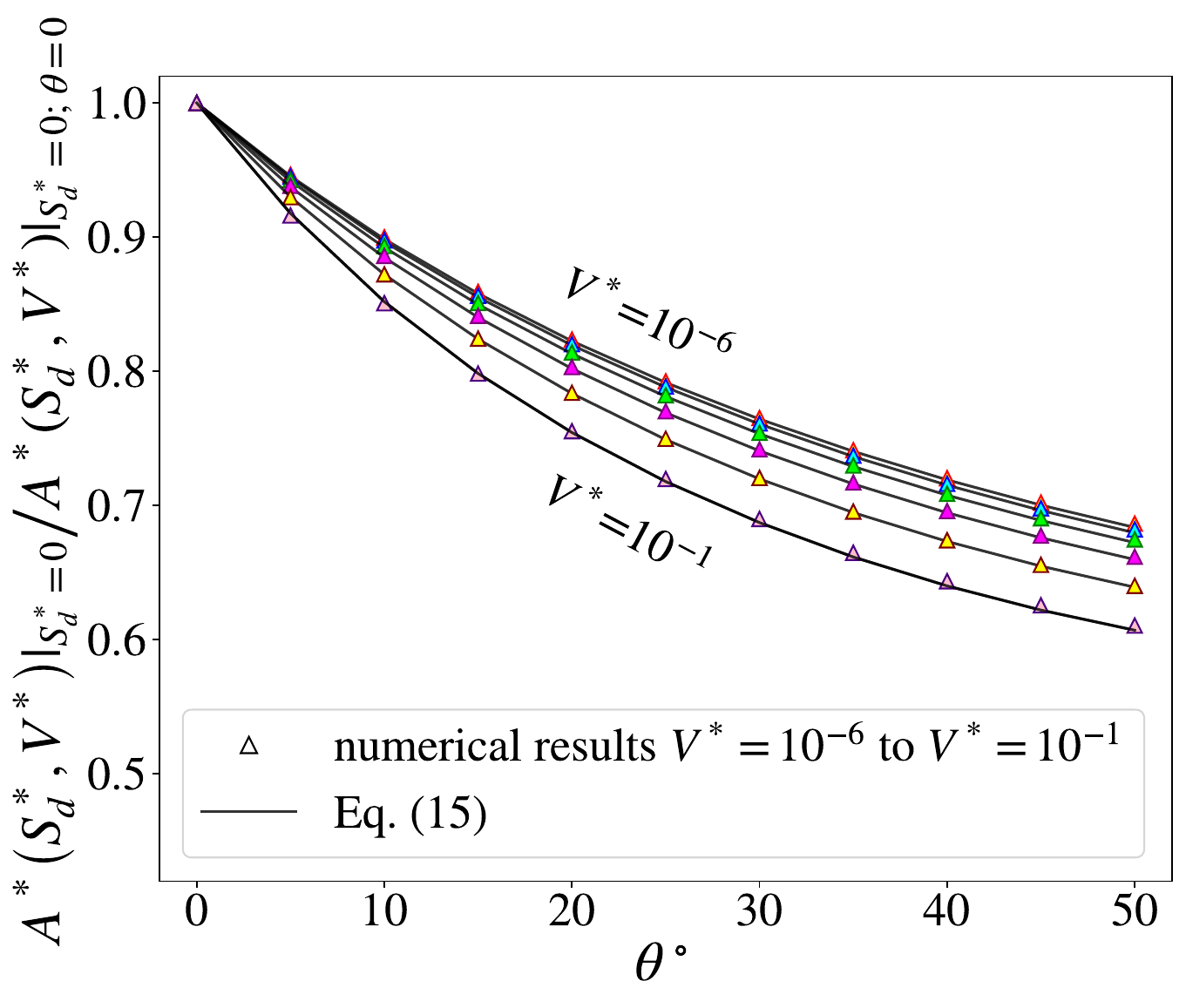}
    \caption{Surface area of a liquid bridge between spheres in contact, as a function of the contact angle for various fixed volumes, $V^*\in\left\{10^{-6}, 10^{-5}, 10^{-4}, 10^{-3}, 10^{-2}, 10^{-1}\right\}$. Symbols: numerical results; lines: fit formula, Eq. \eqref{eq Atheta/A0}.}
    \label{fig Area/ContactAngle}
\end{figure}

Both the decrease in the half-filling angle and the mean curvature with increasing contact angle cause the curved surface of the liquid bridge to assume a nearly cylindrical shape for large $\theta$ (see \autoref{fig liquid bridge profile}),  which in turn reduces the accessible surface of the liquid bridge. Figure \ref{fig Area/ContactAngle} 
shows the surface area of the liquid bridge spanning between contacting spheres as a function of the contact angle, $\left.A^*(S_d^*,V^*)\right|_{S^*_d=0}$, as obtained from the numerical solution of Eq. \eqref{eq:a}, for several values of the volume, $V^*\in\left[ 10^{-6},10^{-1}\right]$. The data are normalized by $\left. A^*(S_d^*,V^*)\right|_{S_d^*=0;\, \theta=0}$. The solid lines show the function
\begin{equation}
    \frac{\left.A^*(S_d^*,V^*)\right|_{S_d^*=0}  }{\left.A^*(S_d^*,V^*)\right|_{S_d^*=0;\, \theta=0}}=\frac{1}{1+a_\theta \theta+b_\theta \theta^2}
    \label{eq Atheta/A0}
\end{equation}
with the fit parameters
\begin{equation}
\begin{split}
    a_\theta &=0.5615{{V^*}^{-0.00836}}+0.7813{{V^*}^{0.2046}}\\
    b_\theta &=-0.09168{{V^*}^{-0.02997}}-0.4159{{V^*}^{0.1933}}.
\end{split}
    \label{eq ab_teta}
\end{equation}

\subsubsection{Surface area as a function of the volume: General case, $S^*_d\ne 0$, $\theta\ne 0$}
\label{sec:A}
\begin{figure}[!htb]
    \centering
    \includegraphics[width=10cm]{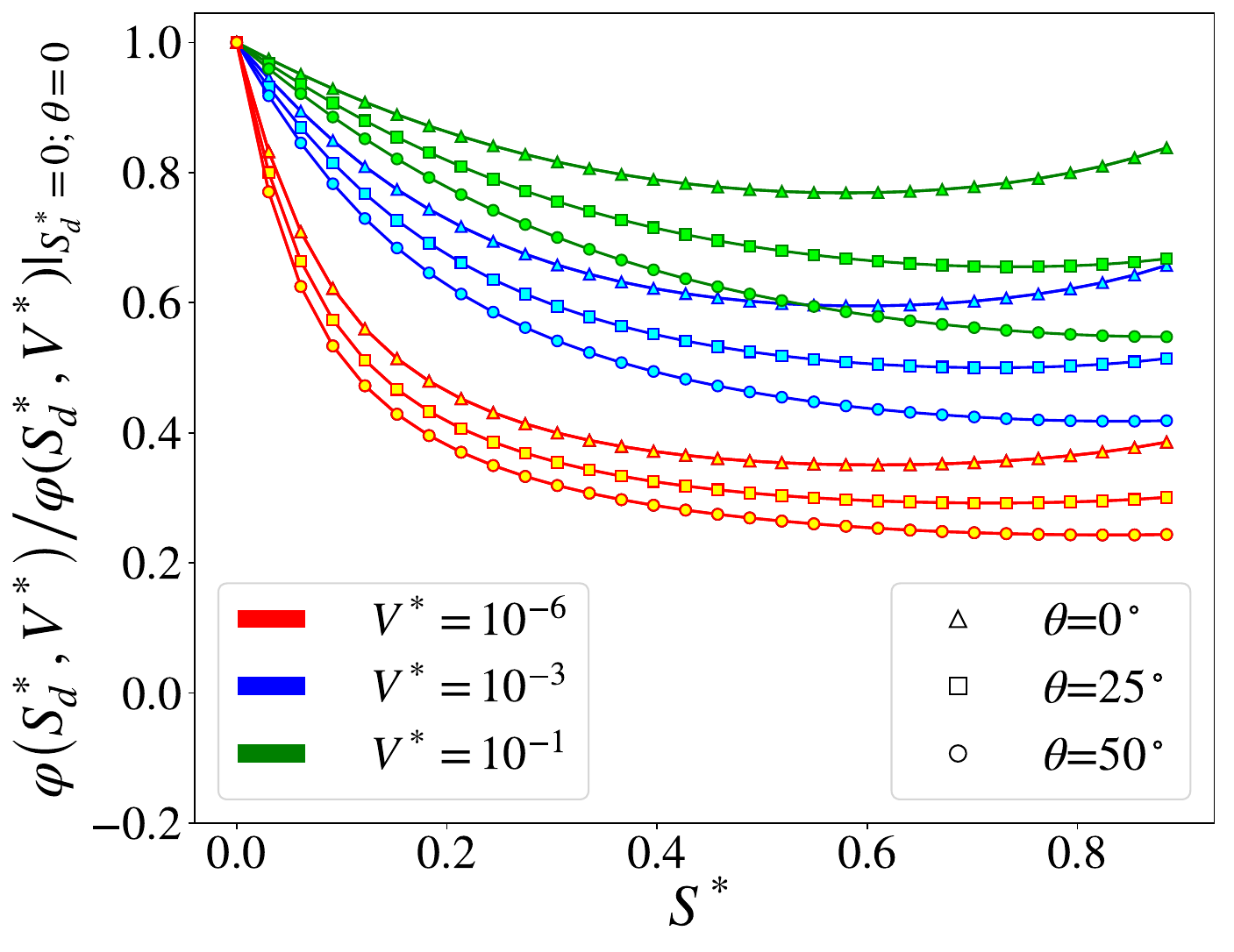}
    \caption{Half-filling angle of a liquid bridge between spheres at distance $S^*$ for volume $V^*\in\left\{10^{-6}, 10^{-3}, 10^{-1}\right\}$ and contact angle $\theta\in\left\{0^\circ, 25^\circ, 50^\circ\right\}$. The data are normalized by $\varphi(S_d^*,V^*)|_{S^*_d=0;\, \theta=0}$. Colors are numerical results for different volumes, and symbols show different contact angles  
    .}
    \label{fig Phi/S}
\end{figure}

\begin{figure}[!htb]
    \centering
    \includegraphics[width=16.5cm]{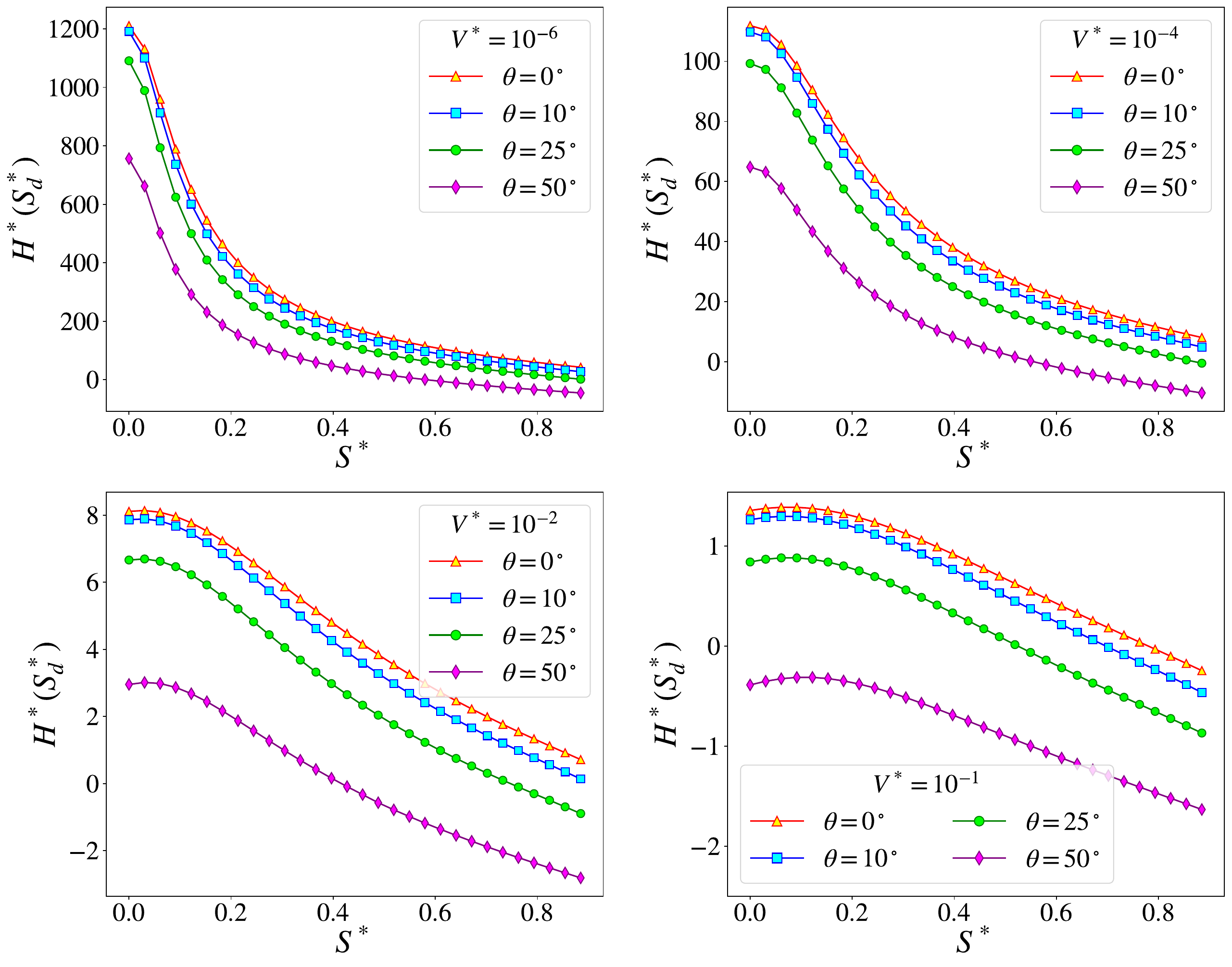}
    \caption{Mean curvature of a liquid bridge as a function of the distance at constant volume, $H^*(S^*)$, for $V^*\in\left\{10^{-6}, 10^{-4}, 10^{-2}, 10^{-1}\right\}$  and contact angle $\theta\in\left\{0, 10^\circ, 25^\circ, 50^\circ\right\}$}
    \label{fig H/S}
\end{figure}

\begin{figure}[!htb]
    \centering
    \includegraphics[width=10cm]{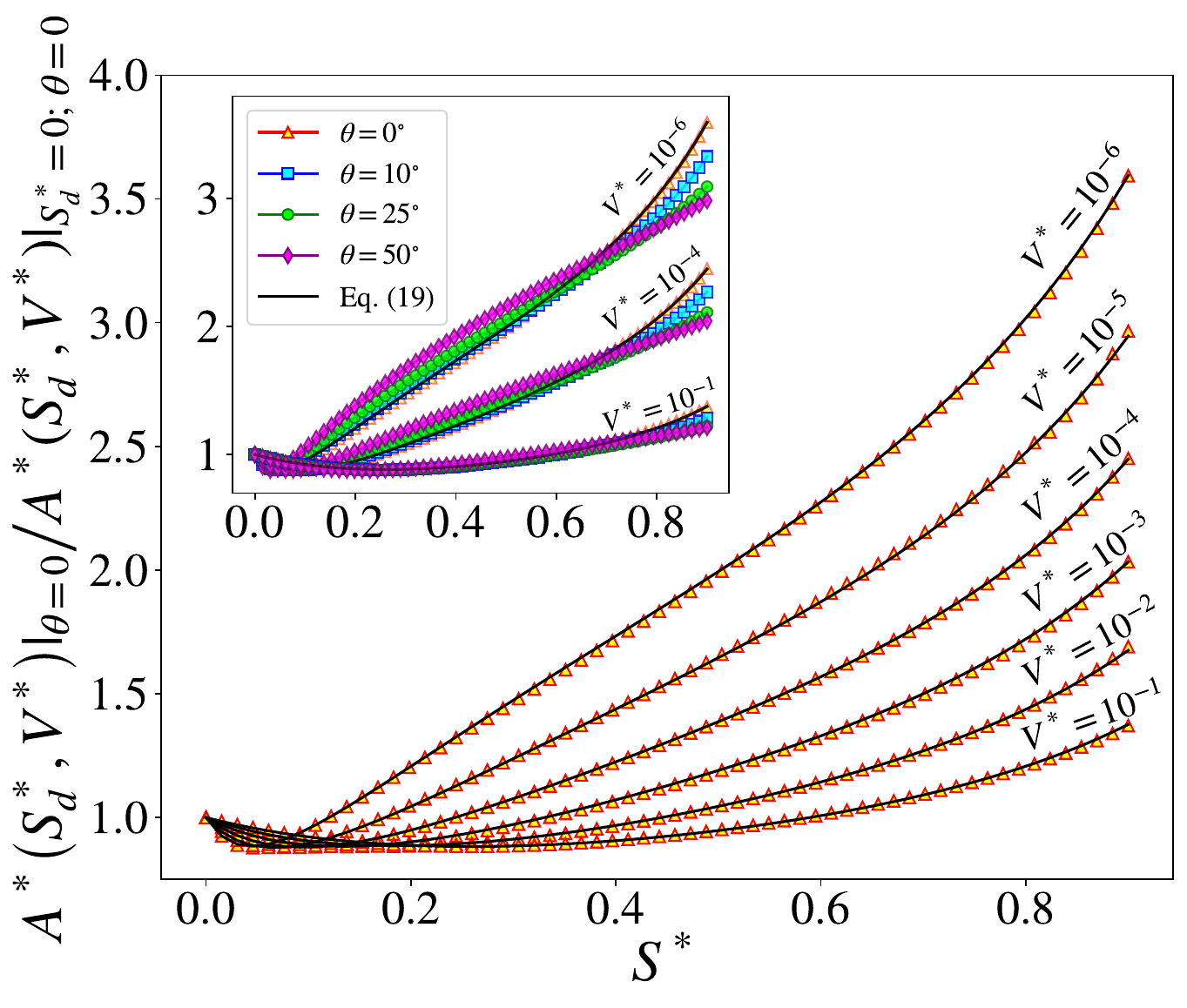}
    \caption{Accessible surface of a liquid bridge spanning between spheres as a function of the distance $S^*$. Symbols: numerical data for constant volume $V^*\in \left\{10^{-6},10^{-5},10^{-4},10^{-3},10^{-2},10^{-1}\right\}$ and contact angle $\theta=0$. Lines: fit formula, Eq. \eqref{eq As/A0}. Inset: same, but for $V^*\in \left\{10^{-6},10^{-4},10^{-1}\right\}$ and $\theta\in\left\{0, 10^\circ, 25^\circ, 50^\circ\right\}$
    }
    \label{fig AsA0/S}
\end{figure}

\begin{figure}[!htb]
    \centering
    \includegraphics[width=10cm]{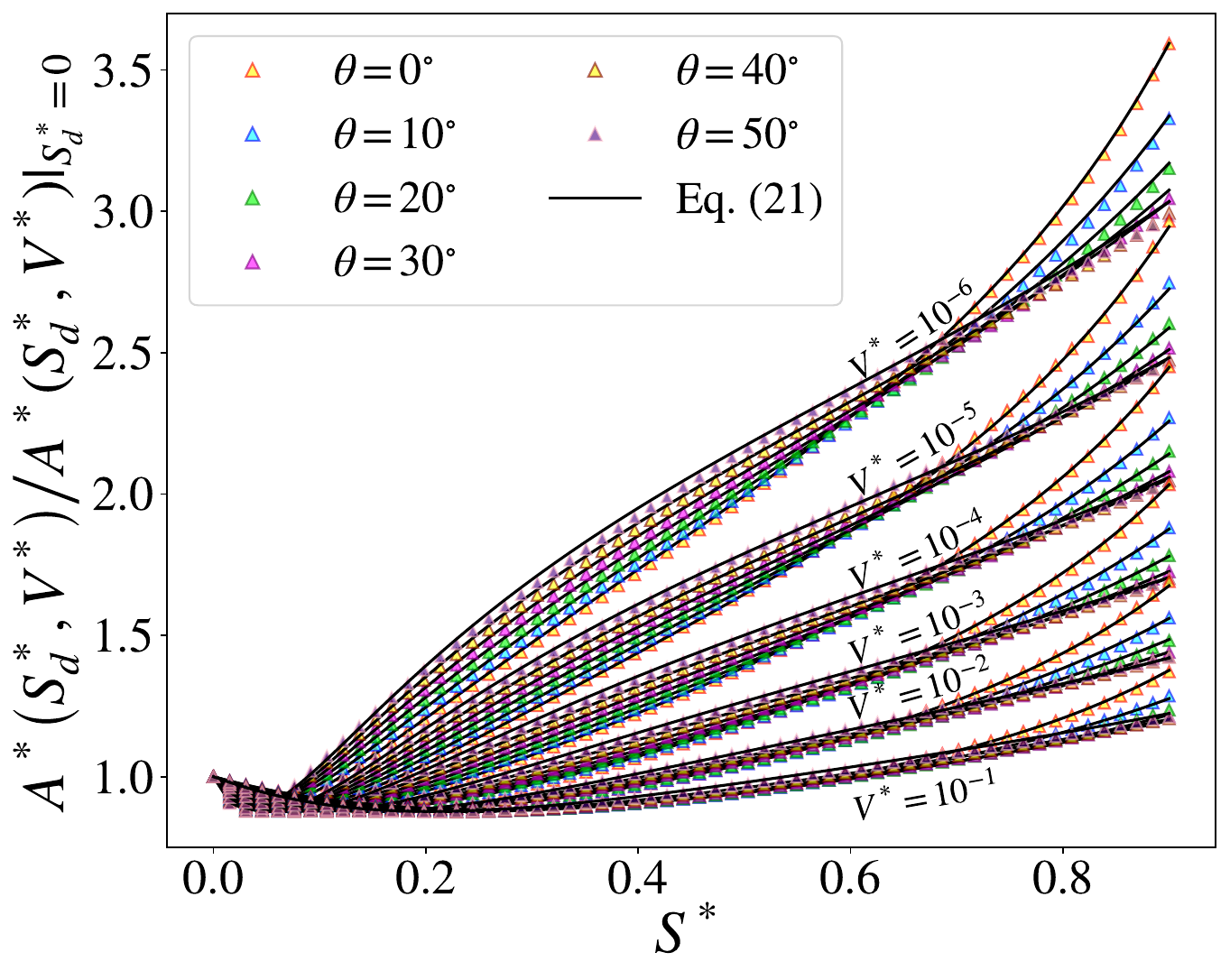}
    \caption{Numerical results for capillary bridges of different volume, $V^*\in\left\{10^{-6},\dots\,10^{-1}\right\}$, and different contact angle, $\theta\in\{0\,\dots\,50^o\}$ as a function of separation distance. Symbols show the numerical integration of Eq. \eqref{eq:a}, and lines show Eq. \eqref{eq A/Atheta}.
    }
    \label{fig AThetaA0/S}
\end{figure}

The critical distance between the surfaces of the spheres when the liquid bridge ruptures is well described by \cite{Willett2000}
\begin{equation}
    S^*_\text{crit}=\left(1+\frac{\theta}{2}\right)\left({V^*}^{1/3}+\frac{{V^*}^{2/3}}{10}\right)\,,
    \label{eq Scrit}
\end{equation}
with $\theta$  in radians, thus, we consider $S^*_d\in\left[0, S^*_\text{crit}\right]$. For fixed volume, we scale the distance
\begin{equation}
    S^*\equiv \frac{S^*_d}{S^*_\text{crit}}\,,
\end{equation}
with $S^*\in[0,1]$. The numerical solution of the Young-Laplace equation for constant liquid bridge volume is presented in Figs. \ref{fig Phi/S}-\ref{fig AThetaA0/S}.

Figure \ref{fig Phi/S} shows the half-filling angle as a function of the separation distance and the volume, $\varphi(S_d^*,V^*)$, normalized by the value for spheres in contact $\left.\varphi(S_d^*,V^*)\right|_{S^*_d=0;\, \theta=0}$.
Data are shown for $S^*\in[0,0.9]$. The numerics become increasingly unstable beyond this interval, next to the rupture distance. For large contact angle values, $\theta=50^\circ$, $\varphi(S_d^*,V^*)$ is a decaying function, while for smaller contact angles, the function reveals a minimum.

Figure \ref{fig H/S} 
shows the mean curvature as a function of the distance at constant volume, $H^*(S_d^*)$, for $V^*\in\left\{10^{-6}, 10^{-4}, 10^{-2}, 10^{-1}\right\}$  and contact angle $\theta\in\left\{0, 10^\circ, 25^\circ, 50^\circ\right\}$.  The function is strictly decaying, except for large volume, $V^*=0.1$, where we see a maximum. This result contrasts previous studies \cite{lian1993, Lian2016}, where a monotonic decrease was reported independent of the volume. 

Our main result is shown in \autoref{fig AsA0/S}.
It shows the accessible surface of a liquid bridge as a function of the distance, normalized by the surface area for spheres in contact, $\left. A^*\left(S^*\right)\right/ \left.A^*\right|_{S^*_d=0;\, \theta=0}$, for constant volume $V^*\in \left\{10^{-6},10^{-5},10^{-4},10^{-3},10^{-2},10^{-1}\right\}$ and contact angle $\theta\in\left\{0, 10^\circ, 25^\circ, 50^\circ\right\}$. The symbols show the numerical data, the lines show the fit formula, Eq. \eqref{eq As/A0}.  Surprisingly, $A^*(S^*)$ is not monotonous but has a minimum. Thus, the smallest surface area is not achieved when the spheres are in contact but at some distance.

We approximate the numerical results by
\begin{equation}
    \frac{\left.A^*(S_d^*,V^*)\right|_{\theta=0}}{\left.A^*(S_d^*,V^*)\right|_{S_d^*=0;\, \theta=0}}
    =\frac{a_s {S^*}^3-b_s {S^*}^2 - c_s {S^*} - 0.01}{a_s {S^*}^4-d_s {S^*}^2 - a_s {S^*} - 0.01}
    \label{eq As/A0}
\end{equation}
with
\begin{equation}
    \begin{split}
         a_s & =0.00539{V^*}^{-0.2332}+0.00029{V^*}^{-0.0971}\\
         b_s & =0.000057{V^*}^{-0.5419}+0.01318{V^*}^{-0.2780}\\
         c_s & =-0.003518{V^*}^{0.0237}+0.00123{V^*}^{-0.2918}\\
         d_s & =-0.003147{V^*}^{-0.0244}+0.00475{V^*}^{-0.2453}
         \label{eq a_s-d_s}
    \end{split}   
\end{equation}
The effect of the contact angle on $A^*$ is illustrated in the inset to \autoref {fig AsA0/S},  for $V^*\in \left\{10^{-6},10^{-4},10^{-1}\right\}$. For all values of $\theta$,  the deviation from the  curves for $\theta=0$ is small and can be covered by a mild modification  of Eq. \eqref{eq As/A0}: 
\begin{equation}
    \frac{A^*\left(S_d^*,V^*\right)}{\left.A^*(S_d^*,V^*)\right|_{S_d^*=0}}
    =\frac{a_s {S^*}^3- e_\theta f_\theta (1+\theta) b_s {S^*}^2 - c_s {S^*} - 0.01}
    {a_s {S^*}^4-f_\theta (1+\theta) d_s {S^*}^2 - e_\theta g_\theta a_s {S^*} - 0.01}
    \label{eq A/Atheta}
\end{equation}
with
\begin{equation}
\begin{split}
    e_\theta &=\frac{a_e\theta^2+a_e\theta-b_e}{\theta-b_e}\\
    f_\theta &=a_f\theta^2+b_f\theta+1\\
     g_\theta&=a_g\theta^2+b_g\theta+1
    \label{eq g_teta-e_teta}
\end{split}
\end{equation}
and the corresponding coefficients depending solely on the volume,
\begin{equation}
\begin{split}
a_e=&0.0000298({\ln{V^*}})^4+0.00111(\ln{V^*})^3+0.01421(\ln{V^*})^2+\\
& + 0.05821\ln{V^*}+0.2012\\
b_e=&0.0000877(\ln{V^*})^4+0.00305(\ln{V^*})^3+0.03849(\ln{V^*})^2+\\
& 0.2521\ln{V^*}+0.243 \\
a_f=&-0.9185{V^*}^{0.0612}-11.46{V^*}^{0.8370}\\
b_f=&3.078{V^*}^{0.09386}+12.77{V^*}^{0.68}\\
a_g=&-0.9588{V^*}^{0.0607}-7.343{V^*}^{0.818}\\
b_g=&3.119{V^*}^{0.07801}+7.643{V^*}^{0.5406}.
\end{split}
\label{eq a_e-b_g}
\end{equation}
Figure \ref{fig AThetaA0/S} 
compares the final fit formula, Eq. \eqref{eq A/Atheta} with the numerical solution of Eq. \eqref{eq:a}, based on the numerical solution of the Young-Laplace equation \eqref{Young-Laplace}. We see that Eq. \eqref{eq A/Atheta} approximates numerical data up to a good precision for the entire range of arguments. The contact angle, $\theta$, can be considered as a further variable, however, it plays more the r\^ole of a parameter since, in most cases, it stays invariant throughout the experiment.

\subsection{Approximative expression $F^*(S^*_d,V^*)$ for the interaction force}
\subsubsection{Outline}
We develop approximative expressions for the interaction force due to a liquid bridge spanning between spheres as a function of the spheres' distance and the volume of the liquid bridge, $F^*(S^*_d, V^*)$, following a similar route as for the accessible surface area, $A^*(S^*_d, V^*)$, considered in Sec. \ref{sec:ApproxSurface}.

First, we derive an expression for the force under the conditions $S_d^*=0$ and $\theta=0$, $F^*\left.(S^*_d, V^*)\right|_{S_d^*=0;\, \theta=0}$. Subsequently, we release the constraint $\theta=0$ to obtain $\left.F^*\left(S_d^*,V^*\right)\right|_{S_d^*=0}$. Finally, we release the condition $S_d^*=0$ to obtain $F^*(S^*_d,V^*)$.

\subsubsection{Capillary force as a function of the volume for particle separation $S^*_d=0$ and contact angle $\theta=0$}
\label{sec:FSd0Theta0}
\begin{figure}[!htb]
    \centering
    \includegraphics[width=10cm]{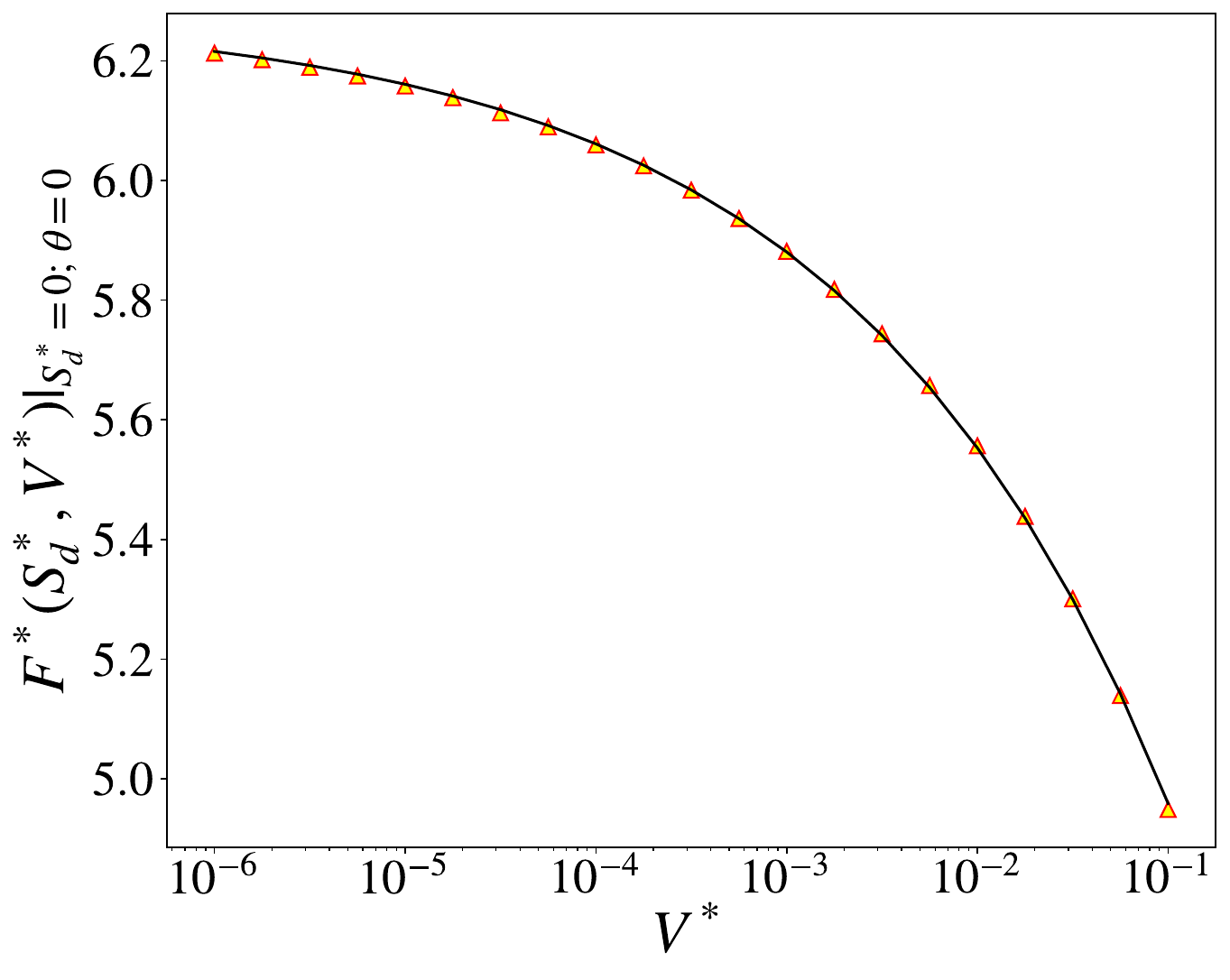}
    \caption{Capillary force between spheres in contact and for zero contact angle as a function of the dimensionless volume, $F^*\left.(S^*_d,  V^*)\right|_{S_d^*=0;\, \theta=0}$. The symbols show the numerical integration of Eq. \eqref{eq:force}, and the line shows the fit given by \eqref{eq F0-V}.
    }
    \label{fig: force at ZeroA}
\end{figure}

For spheres in contact, $S^*_d=0$, the capillary force is a decreasing function of the liquid bridge's volume. Figure \ref{fig: force at ZeroA} shows $F^*\left.(S^*_d, V^*)\right|_{S_d^*=0;\, \theta=0}$ obtained by numerical integration of Eq. \eqref{eq:force}.
The numerical results are well approximated by 
\begin{equation}
    F^*\left.(S^*_d,  V^*)\right|_{S_d^*=0;\, \theta=0} = 2\pi\left(1-0.3823\,{V^*}^{0.2586}\right)\,.
    \label{eq F0-V}
\end{equation}

\subsubsection{Capillary force as a function of the volume for particle separation $S^*_d=0$}
\label{sec:FTheta0}
\begin{figure}[!htb]
    \centering
    \includegraphics[width=10cm]{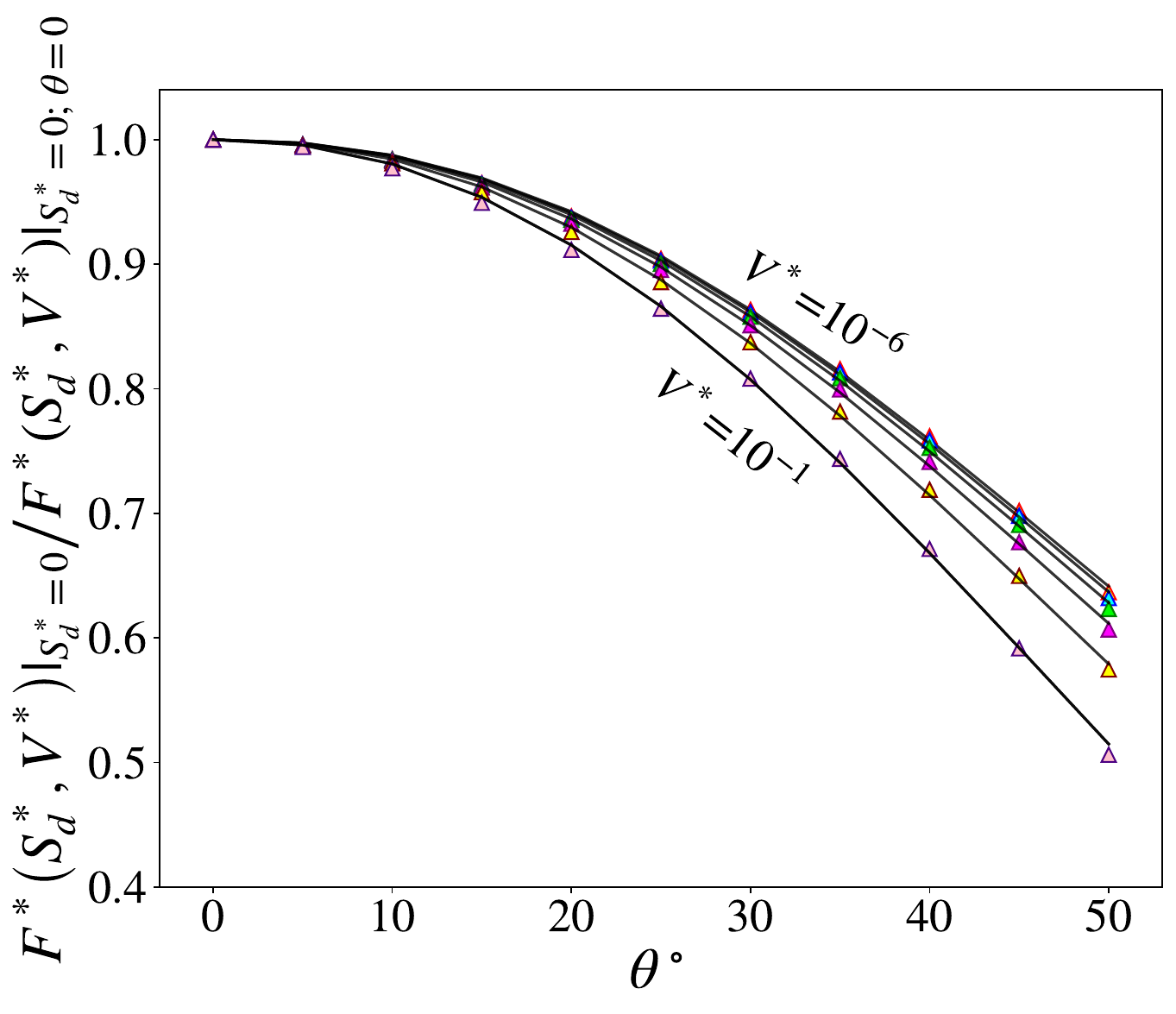}
    \caption{Capillary force between spheres in contact as a function of the contact angle, $F^*\left.(S^*_d,  V^*)\right|_{S_d^*=0}$ for different values of the liquid volume, $V^*\in\left\{10^{-6}, 10^{-5}, 10^{-4}, 10^{-3}, 10^{-2}, 10^{-1}\right\}$.
    The symbols show the numerical integration of Eq. \eqref{eq:force}, and the lines show the fit given by \eqref{eq F0-theta}.
    } 
    \label{fig: force at ZeroB}
\end{figure}

Releasing the constraint $\theta=0$, we obtain $F^*\left.(S^*_d,  V^*)\right|_{S_d^*=0}$ through integration of Eq. \eqref{eq:force}. The normalized forces can be approximated by
\begin{equation}
\frac{F^*\left.(S^*_d,  V^*)\right|_{S_d^*=0}}{F^*\left.(S^*_d,  V^*)\right|_{S_d^*=0;\, \theta=0}}
= \left[1 - a_\theta \sin\left(\theta^{b_\theta}\right)\right]\,,
    \label{eq F0-theta}
\end{equation}
where $\theta$ is the contact angle in radian and $a_{\theta}$ and $b_{\theta}$ are functions of volume:
\begin{equation}
\begin{split}
    a_{\theta} &= 0.4158\,V^{*0.2835} +0.6474\\
    b_{\theta} &= -0.2087\,V^{*0.3113} +2.267\,.
\end{split}
\label{eq f ab-theta}
\end{equation}
Figure \ref{fig: force at ZeroB} shows the interaction force, $F^*\left.(S^*_d,  V^*)\right|_{S_d^*=0}$, as a function of the contact angle, $\theta$, as obtained by numerical integration of Eq. \eqref{eq:force}, in comparison with Eq. \eqref{eq F0-theta} for different values of the liquid volume, $V^*\in\left\{10^{-6}, 10^{-5},\, \dots\,, 10^{-1}\right\}$. The data are normalized by $\left.F^*(S_d^*,V^*)\right|_{S_d^*=0; \theta=0}$. 

This decreasing trend contradicts the simplifying analytical assumptions of the liquid bridge profile, e.g., Derjaguin's equation \cite{Willett2000}, which predicts a constant dimensionless capillary force of $2\pi$ at zero separation \cite{Lian2016}.

\subsubsection{Capillary force as a function of the volume: General case, $S^*_d\ne 0$, $\theta\ne 0$}
\label{sec:Fgeneral}

\begin{figure}[!htb]
    \centering
    \includegraphics[width=10cm]{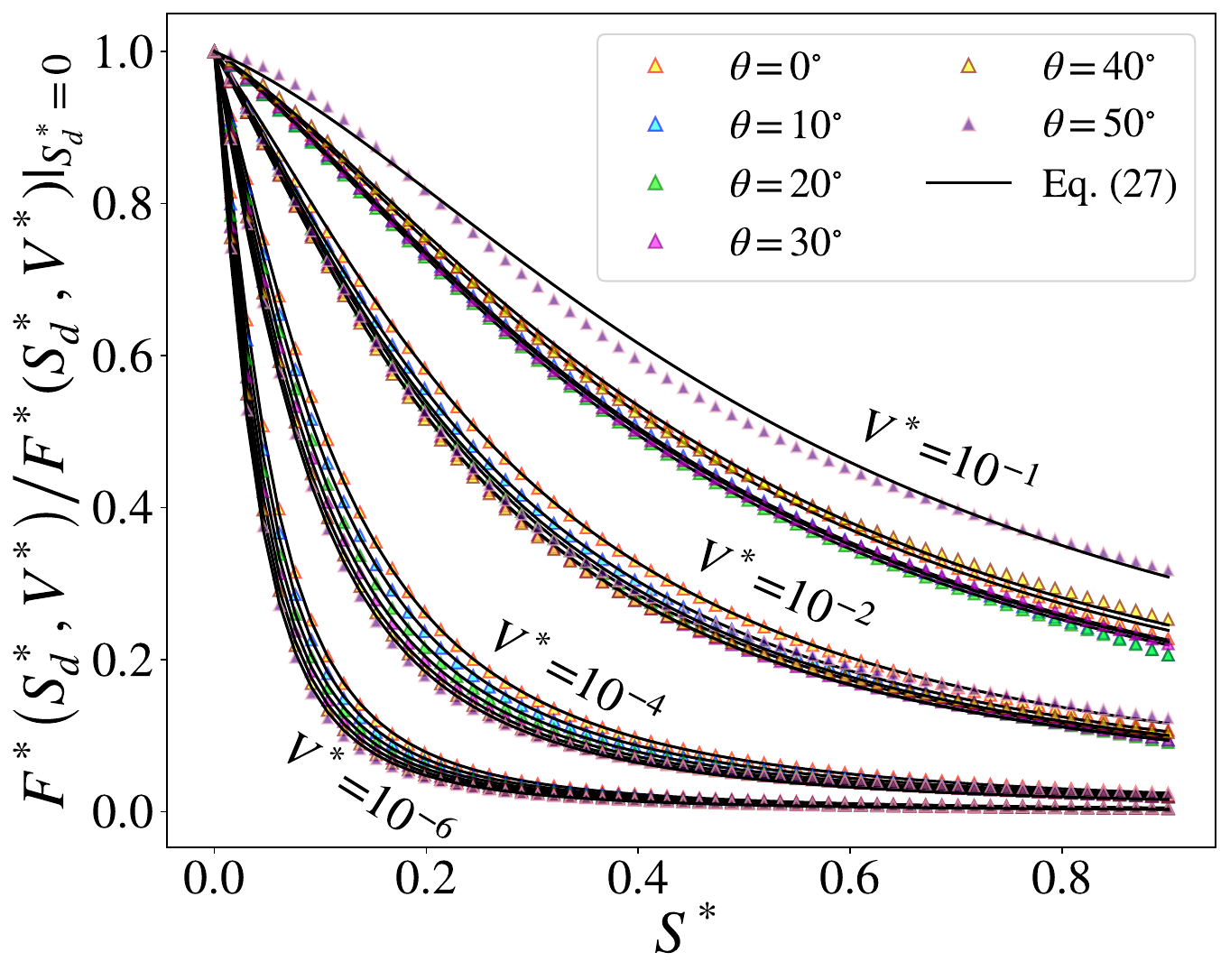}
    \caption{Capillary force due to a liquid bridge spanning between two spheres as a function of the normalized distance $S^*$, for a range of volumes, $V^*\in\left\{10^{-6}, 10^{-4}, 10^{-2}, 10^{-1}\right\}$, and contact angle, $\theta\in\{0, 10^o, 20^o, 30^o, 40^o, 50^o\}$. Symbols: numerical integration of Eq. \eqref{eq:force}, Solid lines: the approximate expression Eq. \eqref{eq F-S}.
}
    \label{fig:F_S}
\end{figure}

We generalize the previous result by releasing the constraint $\theta=0$, that is, we consider spheres that are not in mechanical contact. As in Sec. \ref{sec:A}, we introduce the measure $S^*$ for the distance, defined by Eq. \eqref{eq Scrit}. Figure \ref{fig:F_S} shows the normalized capillary force, $F^*\left(S^*, V^*\right)$, for different volumes, $V^*\in\left\{10^{-6}, 10^{-4}, 10^{-2}, 10^{-1}\right\}$, and contact angles, $\theta\in\{0, 10^o, 20^o, 30^o, 40^o, 50^o\}$ as obtained from the numerical integration of Eq. \eqref{eq:force}.

The numerical results can be approximated by
\begin{equation}
\frac{F^*\left(S^*_d,  V^*\right)}{F^*\left.(S^*_d,  V^*)\right|_{S_d^*=0}}
= 
\frac{1+a_sS^*}{1+ c_{\theta} a_s b_s S^* +c_\theta b_s {S^*}^2} 
    \label{eq F-S}
\end{equation}
where $a_s$ and $b_s$ are pure functions of dimensionless volume:
\begin{equation}
\begin{split}
    {a_{s}} &= -0.3319\,{V^*}^{0.4974}+0.6717\,{V^*}^{0.1995}\\
    {b_{s}} &= 13.84\,{V^*}^{-0.3909}-12.11\,{V^*}^{-0.3945}\,.
\end{split}
    \label{eq f ab-s}
\end{equation}
The effect of the contact angle on the capillary force is taken into account by
\begin{equation}
    {c_{\theta}} = a_c\, \theta^3 + b_c\, \theta + 1
    \label{eq f c-theta}
\end{equation}
with
\begin{equation}
\begin{split}
    {a_{c}} &= -0.007815\,(\ln{V^*})^2-0.2105 \,\ln{V^*}-1.426\\
    {b_{c}} &=-1.78\,{V^*}^{0.8351}+0.6669\,{V^*}^{-0.0139}.
    \label{eq f ab-c}
\end{split}
\end{equation}
The approximate equation, Eq. \eqref{eq F-S}, is shown in \autoref{fig:F_S} with solid lines to compare with the direct numerical solutions (symbols).

\section{Conclusion}
We present approximative expressions for the area of the accessible surface and the capillary force of a liquid bridge spanning between identical spheres as a function of the liquid volume and the distance of the particles. The equations are scaled by the particle radius, therefore, the radius does not appear in the final expressions. The surface area depends, moreover, on the liquid-solid contact angle and the surface tension coefficient. These quantities enter the expression as parameters.

These approximate expressions can be used in large-scale numerical simulations of drying suspensions where millions of capillary forces lead to macroscopic plastic deformations of materials, such as fragmentation. Here, the dynamics is determined by the local evaporation rate governed by the accessible surface of liquid bridges and by the capillary forces exerted on the particles by the liquid bridges. Instead of solving the Young-Laplace equation numerically in each time step, using the approximate expressions can accelerate the simulation considerably. 

\appendix
\section{Numerical solution of the Young-Laplace equation}
\label{sec:NumMethod}
To solve the Young-Laplace equation \eqref{Young-Laplace} with the boundary conditions \eqref{boundary conditions}, we use a scheme similar to \citeauthor{Lian2016}\cite{Lian2016}. Our method differs from their work in that we maintain a constant volume for the liquid bridge across the steps, whereas they keep the half-filling angle constant. 

We write the second-order Young-Laplace equation as a system of two first-order equations and corresponding boundary conditions:
\begin{align}
\frac{\text{d}y^*}{\text{d}x^*} &= z^*(x) & \text{;~} & \left.\frac{\text{d}y^*}{\text{d}x^*}\right|_{-x_c^*} = \sin\varphi\\
\frac{\text{d}z^*}{\text{d}{x^*}} &= 2H^*\left[1+{{z^*(x)}}^2\right]^{3/2}+\frac{1+{z^*(x)}^2}{y^*(x)}   & \text{;~} & \left.\frac{\text{d}z^*}{\text{d}x^*}\right|_{-x_c^*} =-\cot{(\varphi+\theta)}.
\end{align}
For the given half-filling angle, $\varphi$, contact angle, $\theta$, and separation distance, $S^*_d$,  the value of $x_c^*$ and the boundary conditions are given by Eqs. \eqref{eq:xc} and \eqref{boundary conditions}, respectively. 

The mean curvature, $H^*$, is unknown and must be determined numerically. To this end, we exploit the toroidal approximation \cite{lian1993}, which approximates the shape of the liquid bridge by a torus. The corresponding mean curvature reads
\begin{equation}
    H^*_\text{tor}=\frac{2\sin{\varphi}-R^*_a-R^*_m}{2R_m\sin{\varphi}}\,.
\end{equation}
The meridional and the azimuthal radii, 
\begin{align}
R^*_m & \equiv\frac{R_m}{R}\equiv\frac{\frac{S^*_d}2+1-\cos{\varphi}}{\cos{(\varphi+\theta)}}\\
R^*_a&\equiv\frac{R_a}{R}\equiv\sin{\varphi}-\frac{\left[1-\sin{(\varphi+\theta)}\right]\left[\frac{S^*_d}2+1-\cos{\varphi}\right]}{\cos{(\varphi+\theta)}}
\end{align}
are shown in Fig. \ref{fig LB}. 

The numerical integration delivers the approximation $y^*_\text{app}(x^*)$ of the function $y^*(x^*)$, and the solution would meet the second boundary condition at $x^*=x^*_{root}$. Therefore, the deviation
\begin{align}
    \left|x^*_\text{root} - x^*_c\right|  \quad  
\end{align}
can be used as an optimization criterion to determine the correct
value of $H^*$. We iterated the computation to obtain the mean
curvature $H^*$ corresponding to $x^*_\text{root}$ with an accuracy of \(|x^*_\text{root} - x^*_c|<10^{-12}\).

\section{Numerical scheme for the derivation of approximate expressions}

In Sec. \ref{ResultsAndDiscussion}, we derived approximate equations under the condition of a constant volume of the liquid bridge. When solving the Young-Laplace equation, Eq. \eqref{Young-Laplace}, the liquid bridge's volume does not enter as a parameter. Therefore, to ensure a particular liquid volume, $V^*$, at a specific separation distance, we employ an iterative approach that relies on the half-filling angle:

Initially, for the case of spheres in contact, $S^*_d=0$ and contact angle $\theta=0$ are considered in Secs. \ref{sec:Sd0_theta0} and \ref{sec:FSd0Theta0}, we use an iterative root-finding method to calculate the half-filling angle $\varphi$ for a liquid bridge with a specified volume $V^*$. We start the iteration with 
\begin{equation}
\varphi_\text{guess}=1.1206 {V^*}^{0.2681}.
\label{eq Lian fit}
\end{equation}
due to the approximation given in Ref. \citenum{Lian2016}. Solving the Young-Laplace equation results in a liquid bridge profile corresponding to the volume $V^*_\text{guess}$. The half-filling angle, $\varphi_\text{guess}$, ist then iteratively optimized until $\left|V^*_\text{guess}- V^*\right|< 10^{-16}$.

In Secs \ref{sec:Sd0} and \ref{sec:FTheta0} where we consider the case $S^*_d=0$; $\theta\ne 0$, for fixed contact angle, $\theta\in (0^\circ$ to $50^\circ)$, we determine $\varphi$ corresponding to a specific volume in the same way, where we use the solution for $S^*_d=0$; $\theta=0$ as initial guess.

For the general case considered in Secs. \ref{sec:A} and \ref{sec:Fgeneral} we determine $\varphi$ in the same way, where the solutions for $S^*_d=0$; $\theta\ne 0$ serve as initial guess.

\bibliography{liquid-bridge-library}

\end{document}